\documentclass[preprint,showpacs,preprintnumbers,amsmath,amssymb,floatfix]{revtex4}

\usepackage{graphicx}

\begin{document}

\preprint{DESY~12--243
\hspace{11.0cm} ISSN 0418-9833}

\title{Inclusive lepton production from heavy-hadron decay in
 $pp$ collisions at the LHC.}

\author{Paolo Bolzoni}
\email{paolo.bolzoni@desy.de}
\author{Gustav Kramer}
\email{gustav.kramer@desy.de}
\affiliation{{II.} Institut f\"ur Theoretische Physik,
Universit\"at Hamburg, Luruper Chaussee 149, 22761 Hamburg, Germany}

\date{\today}

\begin{abstract}
We present predictions for the inclusive production of leptons 
($e^{\pm},\mu^{\pm}$)
originating from charm and bottom-hadrons at the CERN LHC in 
the general-mass variable-flavor-number scheme at next-to-leading order. 
Detailed numerical results are compared to data of the CMS, ATLAS
and ALICE collaborations. 
\end{abstract}

\maketitle

\section{Introduction}

The investigation of heavy flavour (charm or bottom) production in proton-proton
collisions at the LHC (Large Hadron Collider) are important for testing
perturbative QCD calculations in a new energy domain, where very small 
Bjorken-$x$ momentum fractions are expected to be probed. The detection of the 
heavy hadrons containing charm or bottom quarks can be done in several ways, 
either through their non-leptonic weak decays or through their semileptonical 
decays, where electrons or muons are measured inclusively.

Already at the RHIC (Relativistic Heavy Ion Collider), the PHENIX and STAR
Collaboration \cite{1,2} measured the production of muons and electrons from
heavy flavour decays in $pp$ collisions at $\sqrt{S}=0.2$ TeV. These data have 
been compared with perturbative QCD calculations in the FONLL framework 
\cite{3bis,3} and were found in agreement with the measurements within the 
experimental and theoretical uncertainties.

Experimentally the measurement of the semileptonic cross sections suffers from a
large background due to the semileptonic decay of primary light hadrons
including pions and kaons (the main contribution) and other meson and baryon
decays (such as $J/\psi$ and low mass resonances $\eta,~\rho,~\omega$ and 
$\phi$), from secondary leptons produced from secondary light hadron decays, and
from secondary hadrons escaping from material surrounding the tracking 
chambers. All these backgrounds must be subtracted based on Monte Carlo 
simulations using the usual event generators to obtain the semileptonic yield 
due to the decay of heavy hadrons.

The measured cross sections for the inclusive production of leptons from heavy
hadron decays are of two types. In two experiments the production cross section
is measured separately for solely $b$-hadron decays \cite{4, 5}. In the CMS
experiment \cite{4} the $b$-hadron cross section was discriminated by measuring
the muon transverse momentum with respect to the closest jet. In the ALICE
experiment \cite{5} the production cross section of electrons, $(e^++e^-)/2$ 
from semileptonic bottom hadron decays was selected by using the information on
the distance of the secondary decay vertex with respect to the primary vertex. 
Due to their long lifetime bottom hadrons decay at a secondary vertex displaced
in space from the primary collision vertex. In all other measurements of the
semileptonic production which were presented by the ATLAS \cite{6} and ALICE
\cite{7,8,9} collaborations no attempt was made to separate leptons from
charm and bottom hadrons. Therefore these cross sections constitute the sum of
lepton (electrons or muons) production cross sections of all charmed hadrons 
($D^0,D^+,D_s$ and $c$-baryons) and all bottom hadrons ($B^0,B^+,B_s$ and 
$b$-baryons).

At LHC energies the cross sections of charm and bottom production using special
non-leptonic decays of charm and bottom hadrons have been reported. So, ALICE
\cite{10}, ATLAS \cite{11} and LHCb \cite{12} presented cross section data
on $D^0,D^+,D^{*+}$ and $D_s$ production and CMS \cite{13} published 
measurements for $B^+,B^0,B_s$ and $\Lambda_b$ production cross sections. All 
these cross sections have been calculated in the framework of the 
general-mass-variable-flavour-number-scheme (GM-VFNS) \cite{14}, which has been
developed in the last ten years for various processes. In this scheme we 
calculated the inclusive B meson production cross section in $p\bar{p}$ 
collision \cite{15} at $\sqrt{S}=1.96$ TeV and in $pp$ collisions \cite{16}
and found good agreement with the respective data from the CDF run II \cite{17}
and also with the data from the CMS collaboration \cite{13} at the LHC at 
$\sqrt{S}= 7$ TeV. Besides the data based on special non-leptonic decays of B 
mesons mentioned above there exist also several cross section measurements for
$p\bar{p} \to BX$ and $pp \to BX$ followed by the inclusive decays 
$B \to J/\psi X$ and $B \to \psi (2S)X$. These cross sections have been
calculated recently also in the GM-VFNS \cite{18} and compared with
experimental data from CDF \cite{17,19}, CMS \cite{20}, LHCb \cite{21},
ATLAS \cite{22} and ALICE \cite{23} collaborations at the LHC. The agreement 
between the calculated cross sections and the data was satisfactory. In the 
same scheme the inclusive production of charmed mesons $D^0,D^+,D^{*+},D_s$ and 
the charmed baryon $\Lambda_c$ has been calculated and the results for the 
ALICE \cite{10}, ATLAS \cite{11} and LHCb \cite{12} kinematic conditions have 
been presented in ref. \cite{24} and compared to the experimental data in the 
respective presentations of the ALICE \cite{10}, ATLAS \cite{11} and LHCb 
\cite{12} collaborations and for the ALICE data also in \cite{24}.

The GM-VFNS is similar to the zero-mass variable-flavour-number scheme
(ZM-VFNS), in which the heavy quark mass $m$ is neglected in the calculation of the
hard-scattering cross sections. The predictions in the ZM-VFNS are expected to
be reliable only in the region of very large values of the transverse momentum
$p_T$ of the produced heavy hadron since terms of the order of $m^2/p_T^2$ are 
neglected in the hard-scattering cross sections. In the GM-VFNS these 
$m^2/p_T^2$ terms are retained as they appear in the so-called 
fixed-flavour-number scheme (FFNS) in such a way that by applying appropriate 
subtractions to the FFNS the GM-VFNS approaches the
ZM-VFNS with the usual $\overline{MS}$ prescription in the limit 
$p_T/m \to \infty$. Whereas in the FFNS the gluon and the light 
partons are the only active partons in the initial state and
the heavy quark appears only in the final state, produced in the hard-scattering
process of light partons, in the GM-VFNS as in the ZM-VFNS, due to the
subtraction of the mass-singular contributions in the initial and final state,
the heavy quark appears also in the initial state, and for the final state
appropriate fragmentation functions (FFs) for the transition $heavy-quark \to 
heavy-hadron$ must be introduced which absorb the collinear singular 
contributions in the final state. Such FFs for the transitions $b \to B$,
$b,c \to D^0,~D^+$ and $D^{*+}$ and $b,c \to D_s$ and $\Lambda_c$ have been 
extracted in ref. \cite {15,25,26} at NLO in the $\overline{MS}$ factorization
scheme with $n_f=5$ (for $b$) and with $n_F=4$ (for $c$) flavours, consistently
within the GM-VFN framework using data for the scaled energy ($x$) distribution 
$d\sigma/dx$ of $e^+e^- \to BX$ and $e^+e^- \to DX$ etc, respectively, measured 
by the CERN LEP1, SLAC SLC, CESR CLEO and the KEKB Belle collaborations.

The content of this paper is as follows. In Sec. 2 we summarize our input
choices of PDFs and B- and D-meson FFs. In this section we also explain how the
fragmentation of these mesons into leptons has been calculated. In Sec. 3 we
compare the predictions of the GM-VFN scheme with the existing data from the
recent LHC run at $\sqrt{S}=7$ TeV \cite{4,5,6,7,8,9} and $\sqrt{S}=2.76$ 
TeV \cite{8}. We end with a summary in Sec. 4.

\section{Setup, input PDFs and FFs}

The theoretical framework and results of the GM-VFN approach for $p\bar{p}$ 
($pp$) collisions have been previously represented in detail in refs. 
\cite{14, 15}. Here, we describe our choice of input for the numerical analysis
for inclusive lepton ($e$ or $\mu$) production from charm and bottom hadrons. 
We use for the ingoing protons the PDF set CTEQ 6.6 \cite{27} as implemented in
the LHAPDF \cite{28} library. This PDF set was obtained in the framework of a 
general-mass scheme using the input mass values $m_c=1.3$ GeV, and $m_b=4.5$ 
GeV, and for the QCD coupling $\alpha_s^{(5)}(m_Z) = 0.118$. The $c$-and 
$b$-quark PDFs have the starting scales $\mu_0=m_c$ and $\mu_0=m_b$, 
respectively.

The nonperturbative FFs for the transition of $b \to B$ were obtained by a fit 
to $e^+e^-$ annihilation data from the ALEPH  \cite{29}, OPAL \cite{30} and SLD
\cite{31} collaborations and have been presented in \cite{15}. The combined fit
to the three data sets was performed using the NLO scale parameter 
$\Lambda^{(5)}_{\overline{MS}} = 227$ MeV corresponding to 
$\alpha_s^{(5)}(m_Z) = 0.1181$ adopted from ref. \cite{27}. Consistent with the 
chosen PDFs, the starting scale of the $b \to B$ FF was assumed to be 
$\mu_0=m_b$, while the $q,g \to B$ FFs, where $q$ denotes the light quarks 
including the charm quark, were taken to vanish at $\mu_0$. As input we used 
the FFs with a simple power ansatz which gave the best fit to the experimental 
data. The data from OPAL and SLD included all $b$-hadron final
states, i.e. all $B$ mesons, $B^+$, $B^0$ and $B_s$ and $b$-baryons while in 
the ALEPH analysis only final states with identified $B^+$ ($B^-$) and $B^0$ 
($\bar{B^0}$) were taken into account. For the fit in ref. \cite{15} it was 
assumed that the FFs of all $b$-hadrons have the same shape. This will also be 
assumed in this calculation. In addition we assumed that all $b$-hadrons have 
the same branching fractions and decay distributions into leptons as one of 
the $B$ mesons, $B^+$ or $B^0$. The only difference results from the different 
fractions $b \to b-hadron$, which are taken from the Particle Data Group 
\cite{32}. Based on these values the prediction, for example for $B^0$, is 
multiplied by 2.49, corresponding to leptons coming from $B^0,B^+,B_s$ and 
$\Lambda_b$. The bottom mass in the hard scattering cross sections is 
$m_b=4.5$ GeV as it is used in the PDF CTEQ6.6 and in the FFs for $b \to B$ 
etc.

For the transitions $c \to D^0,~D^+$ we employ the FFs determined in \cite{25}.
They are based on fits to the most precise data on D meson production from the 
CLEO Collaboration at CESR \cite{34} and from the Belle Collaboration at KEKB 
\cite{35}. Actually there are several alternative fits presented in \cite{25}. 
Here we use the so-called Global-GM fit, which includes fitting in addition to 
OPAL data \cite{36} together with the CLEO and Belle results. The fits in
\cite{25} are based on the charm mass $m_c=1.5$ GeV, which is slightly larger 
than the one used in the CTEQ6.6 PDF fits. The starting scale for $c \to D$ is 
$\mu_0=m_c$ as it is for the $g,q \to D$ FFs, whereas for the $b \to D$ FF it is
$\mu_0=m_b$.

The subtractions related to renormalization and to the factorization of initial-
and final- state singularities require the introduction of scale parameters 
$\mu_R$, $\mu_I$ and $\mu_F$. We choose the scales to be of order $m_T$, where
$m_T$ is the transverse mass $m_T=\sqrt{p_T^2+m^2}$ and $m=m_b$ for the case 
of bottom quark production and $m=m_c$ for charm quark production. 
For exploiting the freedom in the choice of scales we introduce the scale 
parameters $\xi_i$ (i=R, I, F) by setting $\mu_i=\xi_i~m_T$. To describe the
theoretical uncertainties we vary the values of $\xi_i$ independently by 
factors of two up and down while keeping any ratio of the $\xi_i$ parameters 
smaller than or equal to two. The uncertainties due to scale variation are the 
dominating source of theoretical uncertainties. Therefore PDF related 
uncertainties and variations of the bottom and charm mass are not considered.

The fragmentation of the final state partons $i$ into lepton $l$ ($l=e^{\pm}$ or
$\mu^{\pm}$) is calculated from the convolution
\begin{eqnarray}
 D_{i \to l}(x,\mu_F) = \int^{1}_{x}\frac{dz}{z}D_{i \to B}(\frac{x}{z},
\mu_F)\frac{1}{\Gamma_B}\frac{d\Gamma}{dz}(z,P_B).
\end{eqnarray}
In this formula $D_{i \to B}(x,\mu_F)$ is the nonperturbative FF determined in 
\cite{15} for the transition $i \to B$ and in \cite {25} for $i \to D$ (with 
subscript $B$ replaced by $D$ in (1)), $\Gamma_B$ is
the total $B$ decay width ($\Gamma_B \to \Gamma_D$ in case of $i \to D$) and 
$d\Gamma(z,P_B)/dz$ is the decay spectrum of $B \to l$ or $D \to l$,
respectively. For given lepton transverse momentum $p_T$ and rapidity $y$, $P_B$
is given by $P_B= |\bf{P_B}|$ = $p_T\sqrt{1+sinh^2y}/z$.
The decay distribution $d\Gamma/dk'_L$, where the momentum $k'_L$ is parallel 
to $\bf{P_B}$, is obtained from the decay distribution in the rest
system of the $B$ meson using the formula (3.16) in ref. \cite{37}, where this 
formula was derived for the decay $B \to J/\psi$ instead of $B \to l$. This 
leads to $d\Gamma(z,P_B)/dz$ used in eq.(1) with $z=k'_L/P_B$.

The electron energy spectrum in inclusive $B \to e\nu X$ decays has been 
measured by the BABAR Collaboration \cite{38} in the range $E_e > 0.6$ GeV up 
to its kinematical limit. This spectrum has been measured at the $\Upsilon(4S)$
resonance. The partial branching fraction for $E_e > 0.6$ GeV has been 
determined as 
${\cal B}[B\to e\nu X, E_e>0.6~GeV]=[10.36\pm0.06(stat)\pm0.23(syst)]\%$

Following an ansatz for the momentum spectrum as a function of $p$ in GeV as 
given in ref. \cite{39} we fitted the BABAR spectrum in the B rest system using 
the formula (3.9) in \cite{37} for the transformation of the spectrum in the 
$B$ rest system to the BABAR laboratory frame, where 
$|\bf{P_B}|$ = $0.341$ GeV by the following formula
\begin{eqnarray}
f_B(p)=(-24.23+14711.2\exp[-1.73\ln^2[2.74-1.10p]])
\\ \nonumber
(-41.79+42.78\exp[-0.5(\sqrt{(p-1.27)^2}/1.8)^{8.78}])
\end{eqnarray}
This differs from the fit in \cite{39} which was done in the BABAR 
$\Upsilon(4S)$ laboratory frame. $f_B(p)$ is related to the partial 
semileptonic decay spectrum $d\Gamma/dp$ by $d\Gamma/dp= c f_B(p)$ with
$c=4.27151~10^{-6}$ according to \cite{37}, where $d\Gamma/dp$ has units 
$[ps^{-1}GeV^{-1}]$. The quality of the fit can be seen in Fig. 1 (left figure) 
together with the effect of the boost into the $B$ rest system, which indeed 
is small. This fit yields for the partial branching fractions
${\cal B} (B \to e\nu X,p>0.6~GeV)=11.04\%$ and 
${\cal B} (B \to e\nu X,p>0)=12.35\%$. The latter branching fraction agrees 
approximately with the more recent PDG value
${\cal B} (B \to e\nu X,p>0)=[10.74 \pm 0.16]\%$ \cite{32}.
\begin{figure*}
\includegraphics[width=6cm]{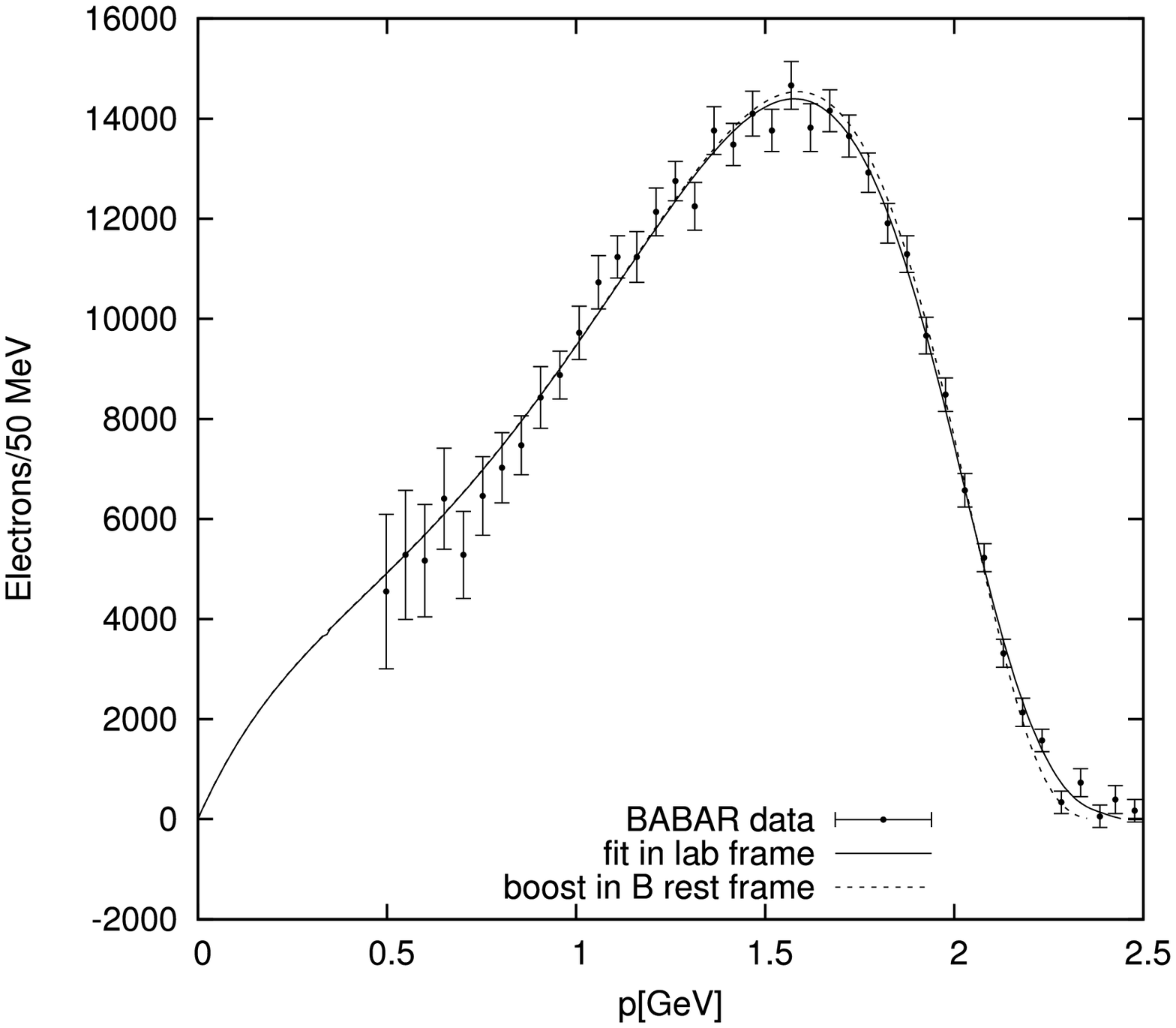}
\includegraphics[width=6cm]{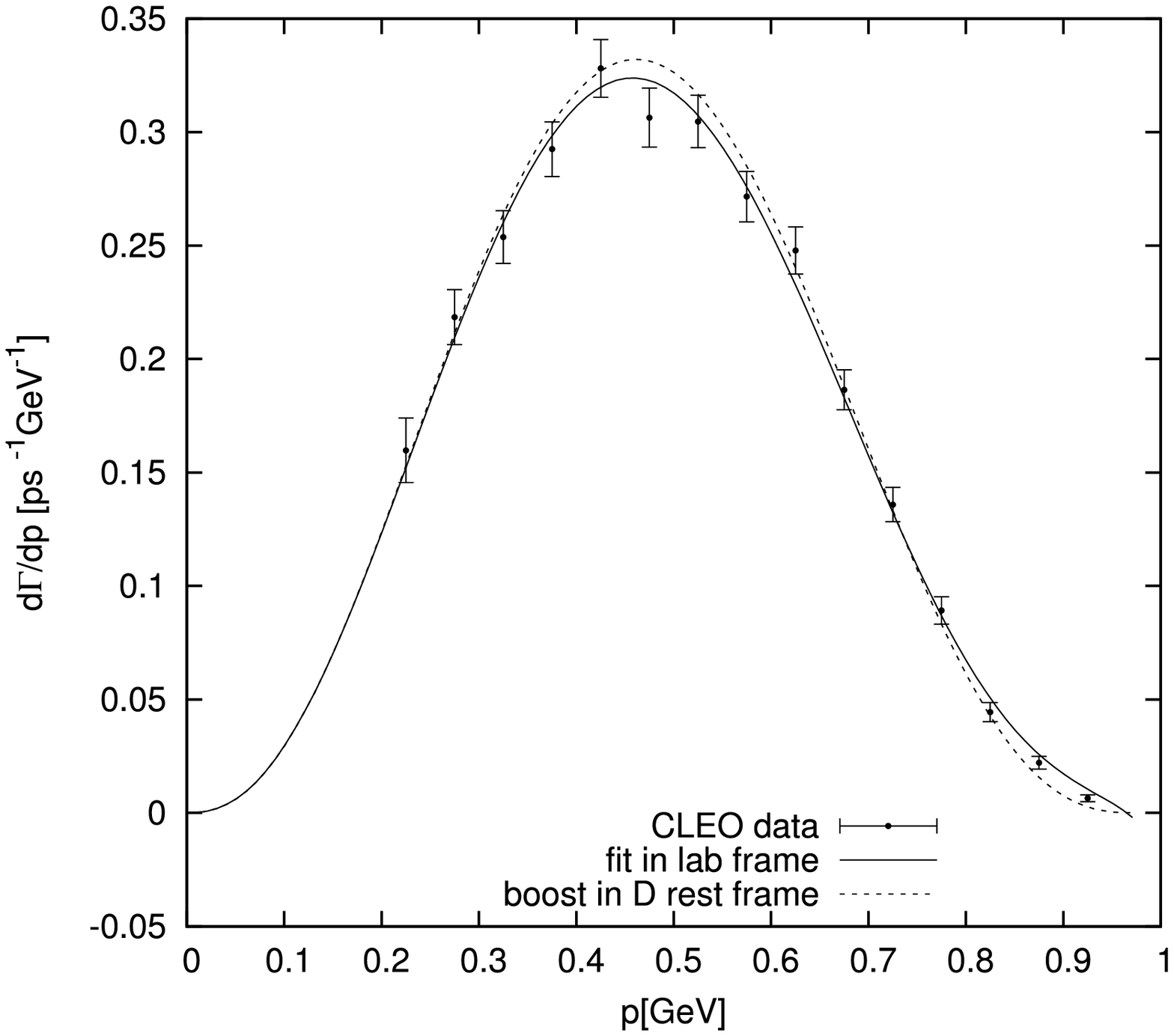}
\caption{\label{fig:fitgraphs1} Fit for inclusive lepton spectrum of $B$ decays
to BABAR (left frame) data in the BABAR laboratory frame together with spectum
boosted to the $B$ rest frame and to inclusive lepton spectrum of $D$ decays to
CLEO (right frame) data in the CLEO laboratoty frame together with the spectrum 
boosted to the $D$ rest frame.}
\end{figure*}

The inclusive electron spectra for the decays $D^+ \to e^+ \nu X$ and 
$D^0 \to e^+\nu X$ have been measured by the CLEO Collaboration \cite{39} on 
the $\psi(3770)$ resonance ($\sqrt{S}=3.73$ GeV) as a function of the electron
momentum $p$. The measured partial branching fractions for $p>0.2$ GeV are 
${\cal B}(D^+\to e^+\nu X, p>0.2~GeV)=[14.97\pm0.19(stat)\pm0.27(syst)]\%$ and
${\cal B} (D^0\to e\nu X,p>0.2GeV)= [5.97\pm0.15(stat)\pm0.10(syst)]\%$ from 
which, on the basis of fits to both spectra by the CLEO Collaboration, the 
following branching fractions 
${\cal B} (D^+ \to e^+\nu X, p>0)=[16.13\pm0.20(stat)\pm0.33(syst)]\%$ and
${\cal B} (D^0 \to e^+\nu X,p>0)=[6.46\pm0.17(stat)\pm0.13(syst)]\%$ result.
The corresponding PDG values are: 
${\cal B} (D^+ \to e^+\nu X)=[16.07\pm0.30]\%$ and
${\cal B} (D^0 \to e^+\nu X)= [6.49\pm0.11]\%$ \cite{32}, which agree very well 
with the CLEO values \cite{40}. The spectra of $D^+$ and $D^0$ as measured
by CLEO coincide very well inside errors in the measured $p$ range $0.2<p<1$ 
GeV. In the CLEO laboratory system the $D^+$ momentum $|\bf{P_D}|$ is $0.243$ 
GeV and the $D^0$ momentum is $0.277$ GeV. These numbers are needed for the 
fits in the $D^+$ and $D^0$ rest systems. The result of the fit for the 
spectrum in the $D$ rest system, for example for the $D^+$ it is
\begin{eqnarray}
 f_D(p) = 17.91(p+0.0034)^{2.66} (0.98-p)^{2.97}
\end{eqnarray}
$f_D(p)$ in eq.(3) is equal to the lepton spectrum $f_D(p)=d\Gamma/dp$ 
in units of $[ps^{-1}GeV^{-1}]$. The quality of the fit to the CLEO data is 
shown in  Fig. \ref{fig:fitgraphs1} in the right frame together with the spectrum in the $D$ rest 
system. As can be seen there the difference between the two spectra is really  
small. The branching ratio for this spectrum is: 
${\cal B} (D^+\to e^++\nu X) = 15.29\%$,
which is sufficiently close to the values from CLEO given above.
With these parametrizations of the
electron spectra in the $B$, respectively $D$, rest system, we calculated the 
lepton spectra $d\Gamma/dx$ in the moving system as function of $k'_L=xP_B$
($xP_D$) where $k'_L$ is the lepton momentum that is parallel to 
$\bf{P_B}$($\bf{P_D}$) using eq.(3.17) in \cite{37}.

To simplify the calculation we applied for $d\Gamma/dx$ the asymptotic formula 
as given in \cite{37}, which is the approximation for $P_B \gg M_B$ 
($P_D  \gg M_D$), where $M_B$ ans $M_D$ are the masses of the $B$ and $D$ meson,
respectively. We calculated $d\Gamma/dx$ for various $P_B$ and found that the 
exact formula differs from the asymptotic formula
by less than $5\%$ for $P_B=10$ GeV. This is easily achieved even for the 
smaller $p_T$ values of the leptons in our applications since $d\Gamma/dx$ is 
peaked at very small $x$, so that the average $x$ is below $0.2$ in the case of
$B$ decays and similarly for $D$ decays. The calculation of the lepton spectra 
is based on the calculation of the cross section $d\sigma/dp_T$ for specific
$D$ meson as reported in Ref.\cite{24}. In the lepton spectra, the leptons
originate from all charmed mesons $D^0,D^+,D_s$ and the charmed baryon $\Lambda_c$.
To make the calculation easier we 
calculated the lepton spectra for just one flavour state, the $D^0$, and 
included the other D mesons, $D^+$, $D_s$ and the contributions  of
charmed baryons, as for example $\Lambda_c$, by an appropriate normalization 
factor. For $D^0$ this normalization factor is $2.294$ and was calculated using
a compilation of charm hadron production fractions by Lohrmann \cite{41} 
(Table 3 of this reference) and the lepton branding fractions for $D^0,D^+,D_s$
and $\Lambda_c$ equal to $0.0646, 0.161, 0.065$ and $0.045$, 
respectively \cite{32}. In case we would have based this normalization factor 
on $D^+$ production, the normalization factor would be slightly different. This 
difference is compensated by the different normalization of $c \to D^+$ and 
the different lepton branching ratio, which result in almost equal cross 
sections $d\sigma/dp_T$ for the production of leptons from charmed hadrons via
$D^0$ or $D^+$.

\section{Results and comparison with LHC data}

\begin{figure*}
\includegraphics[width=6cm]{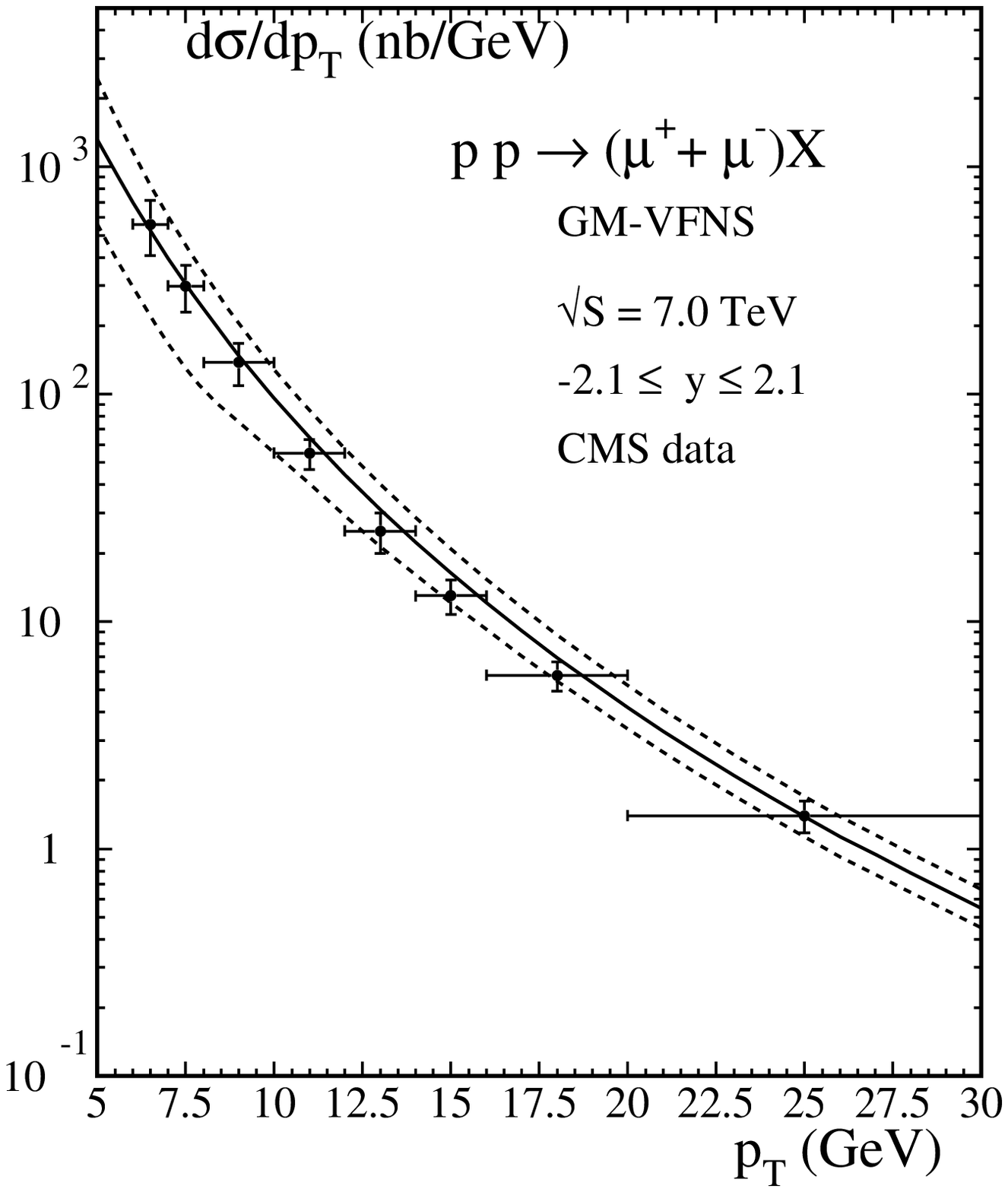}
\includegraphics[width=6cm]{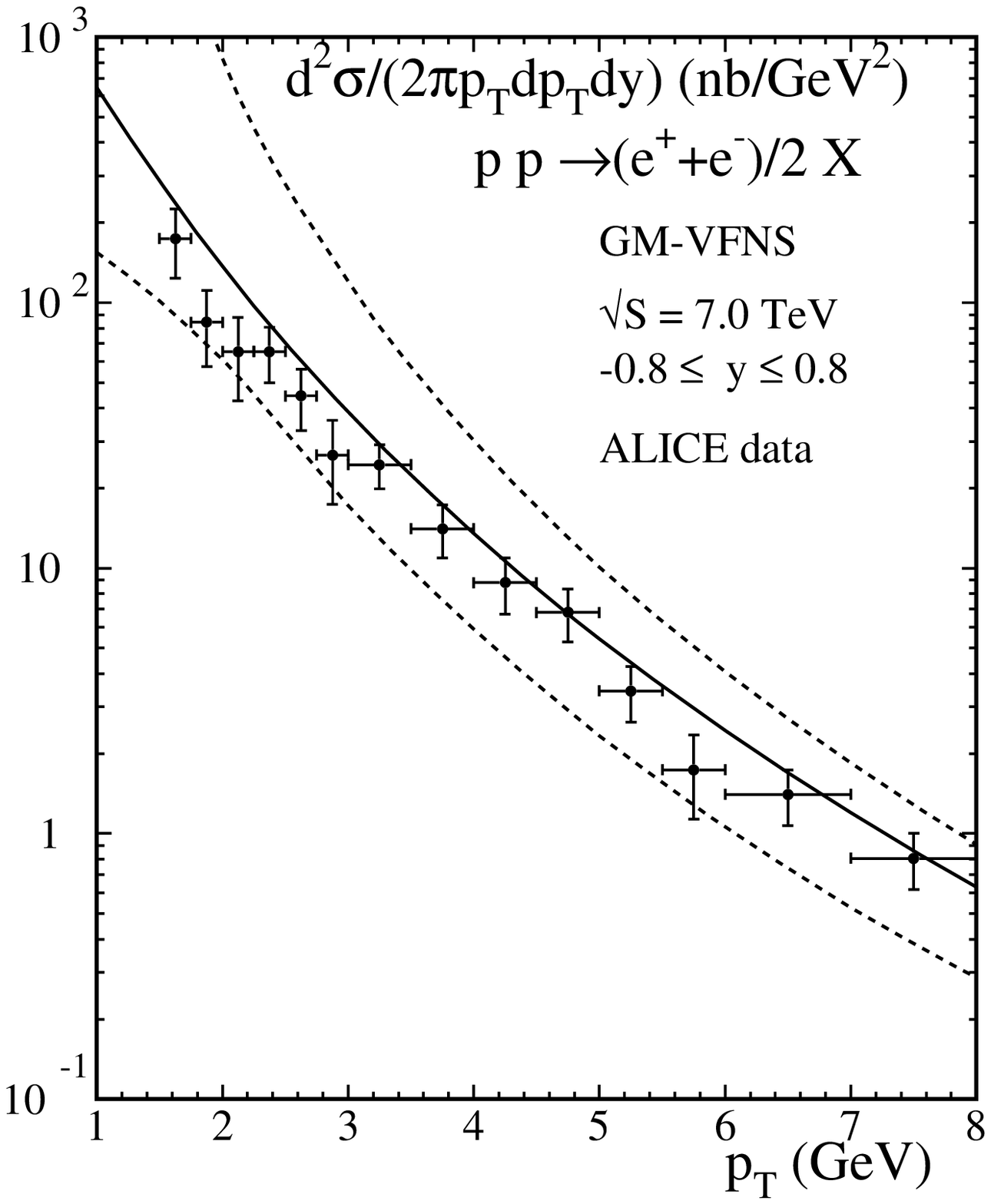}
\caption{\label{fig:fitgraphs2} GM-VFNS predictions for inclusive lepton 
production from bottom hadrons compared to CMS \cite{4} and ALICE \cite{5} data.}
\end{figure*}

In this section we collect our results for the cross sections $d\sigma/dp_T$ as 
a function of $p_T$. for the various LHC experiments and compare them with the 
published data. We start with the results on inclusive b-hadron production with
muons at $\sqrt{S}=7$ TeV by the CMS Collaboration \cite{4} and the ALICE 
Collaboration \cite{5}. The CMS data are for the production of the sum of
$\mu^++\mu^-$ in the range from $6 \leq p_T \leq 30$ GeV, and are integrated 
over the rapidity $y$ in the region $-2.1 \leq y \leq 2.1$. The comparison 
between the GM-VFNS predictions and the experimental data is shown in Fig. \ref{fig:fitgraphs2}. 
The ALICE cross section $d\sigma/dp_T$ \cite{5} for 
the production of electrons and positrons ($(e^++e^-)/2)$ from bottom hadron
decays at $\sqrt{S}=7$ TeV in the rapidity range $-0.8 \leq y \leq 0.8$
is measured for smaller $p_T \leq 8$ GeV. Our prediction in 
comparison with
the ALICE data, is shown also in Fig. \ref{fig:fitgraphs2}, where 
we included the data only for 
$p_T \geq 1.5$ GeV. 
The data agree well with the full curve 
(default scale choice) in both cases, the CMS and the ALICE data.
We remark that in the 
two predictions in Fig. \ref{fig:fitgraphs2} we 
included also the contributions from $b \to D \to lepton$. The FFs needed for 
this contribution were taken from \cite{25} (global fit). These contributions 
are very small, of the order of 1 to 2 $\%$ compared to $c \to D$ contribution,
depending on $p_T$, and are not visible in the logarithmic plots in Fig. \ref{fig:fitgraphs2}. 
The $b \to D$ contributions are also included in all the following predictions.
Note the different form of the cross section in the right frame of Fig. \ref{fig:fitgraphs2}: 
$d^2\sigma/(2\pi dp_T dy)$ is in accordance to the ALICE definition, \emph{i.e.} the 
cross section $d\sigma/dp_T$ is divided by the $y$ bin width $\Delta y=1.6$.
Finally we note that in general at low $p_T$, 
the scale variation of the cross section 
is rather large. Some predictions go down up to 
$p_T=1$ GeV. In this region the scale variation is particularly enhanced and, 
due
to the contribution of b quarks in the initial state and the choice of the 
default scale $\sqrt{p_T^2+m_b^2}$, is not suppressed
as in other approaches as for example MC@NLO \cite{42} in which the small $p_T$ 
predictions are obtained in the FFNS as in the FONLL approach \cite{3,3bis}.
This is an implementation issue in the limit $p_T/m\rightarrow 0$. Further details in the
GM-VFNS framework are discussed in Ref.\cite{24}.

\begin{figure*}
\includegraphics[width=6cm]{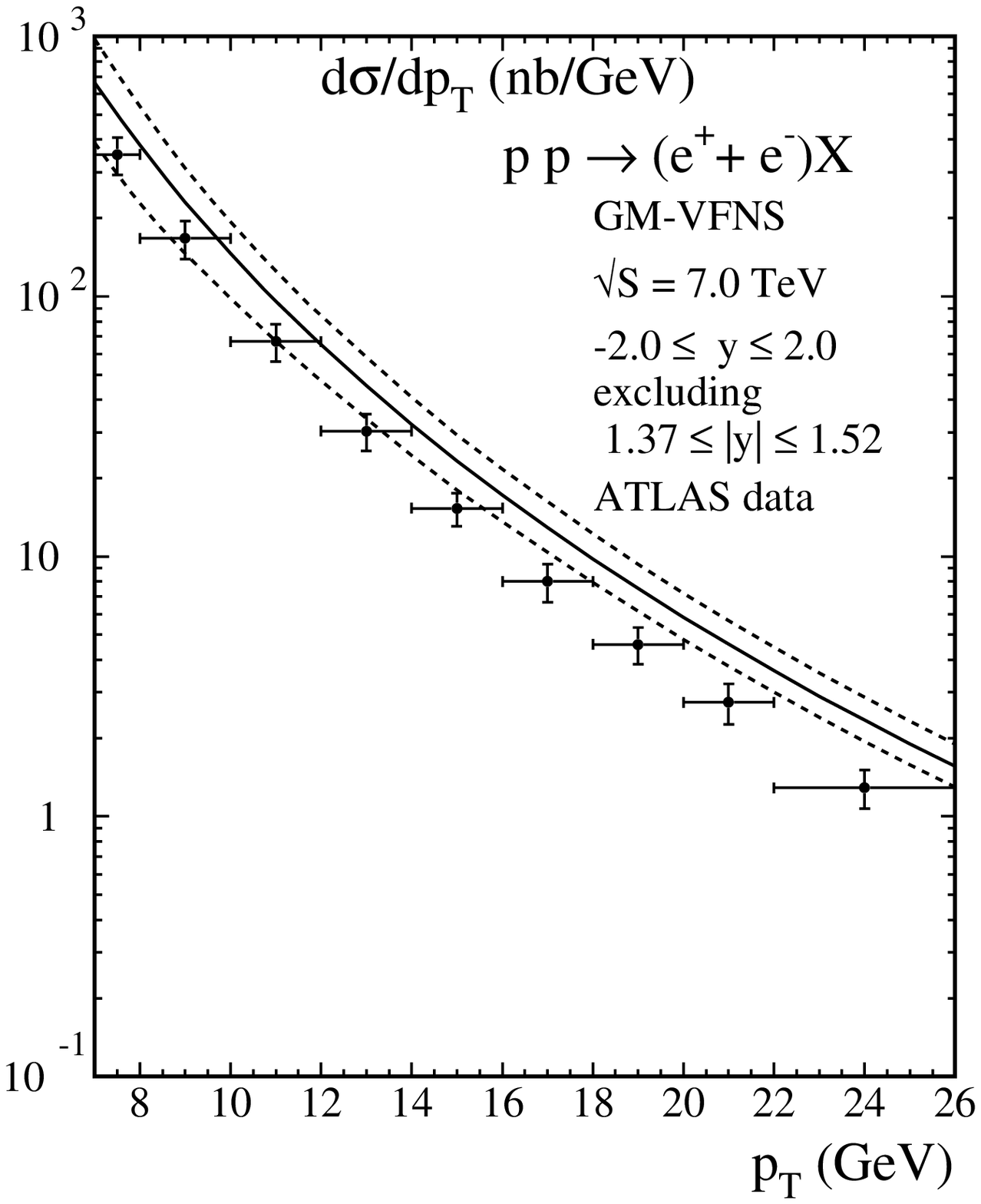}
\includegraphics[width=6cm]{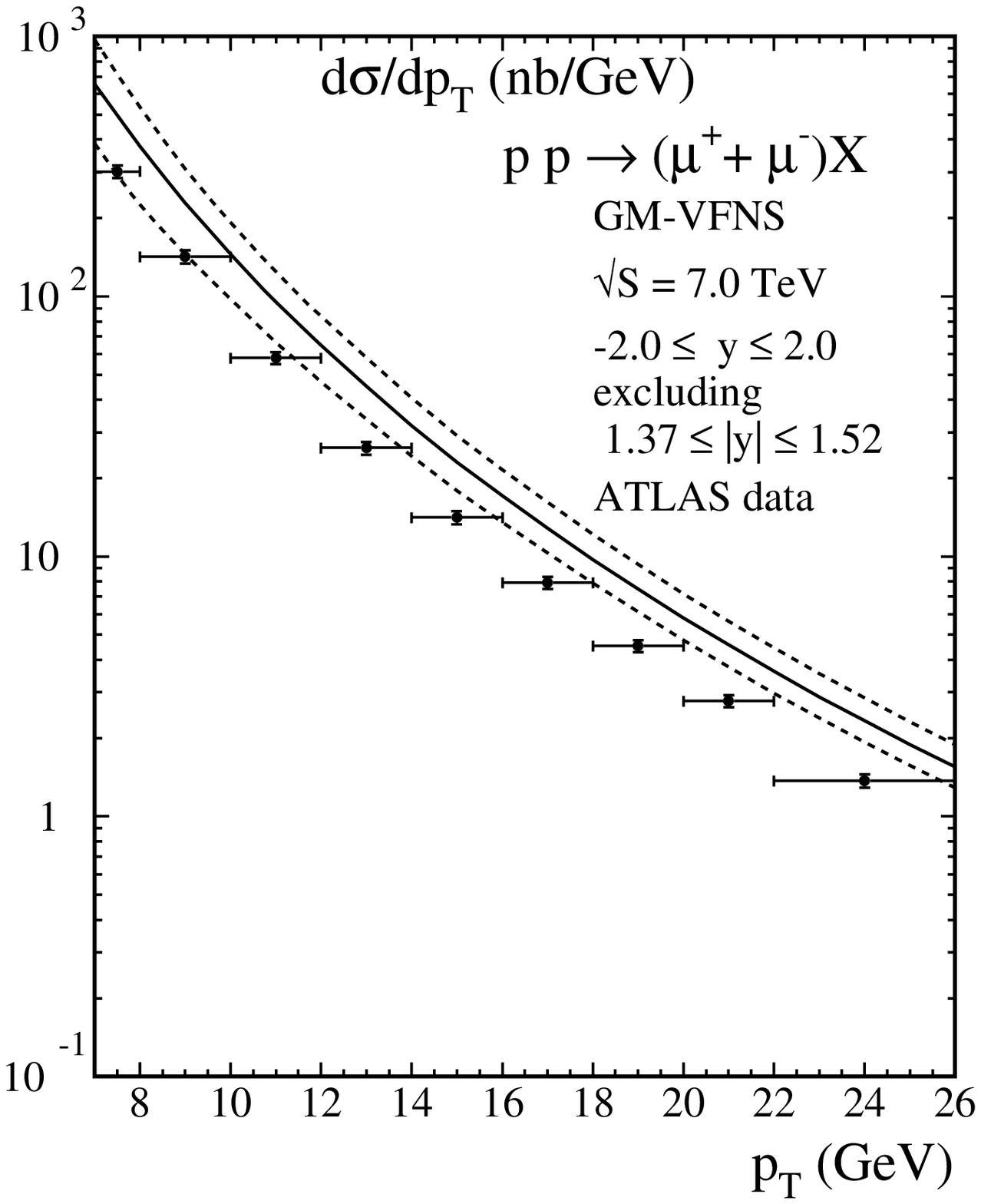}
\includegraphics[width=6cm]{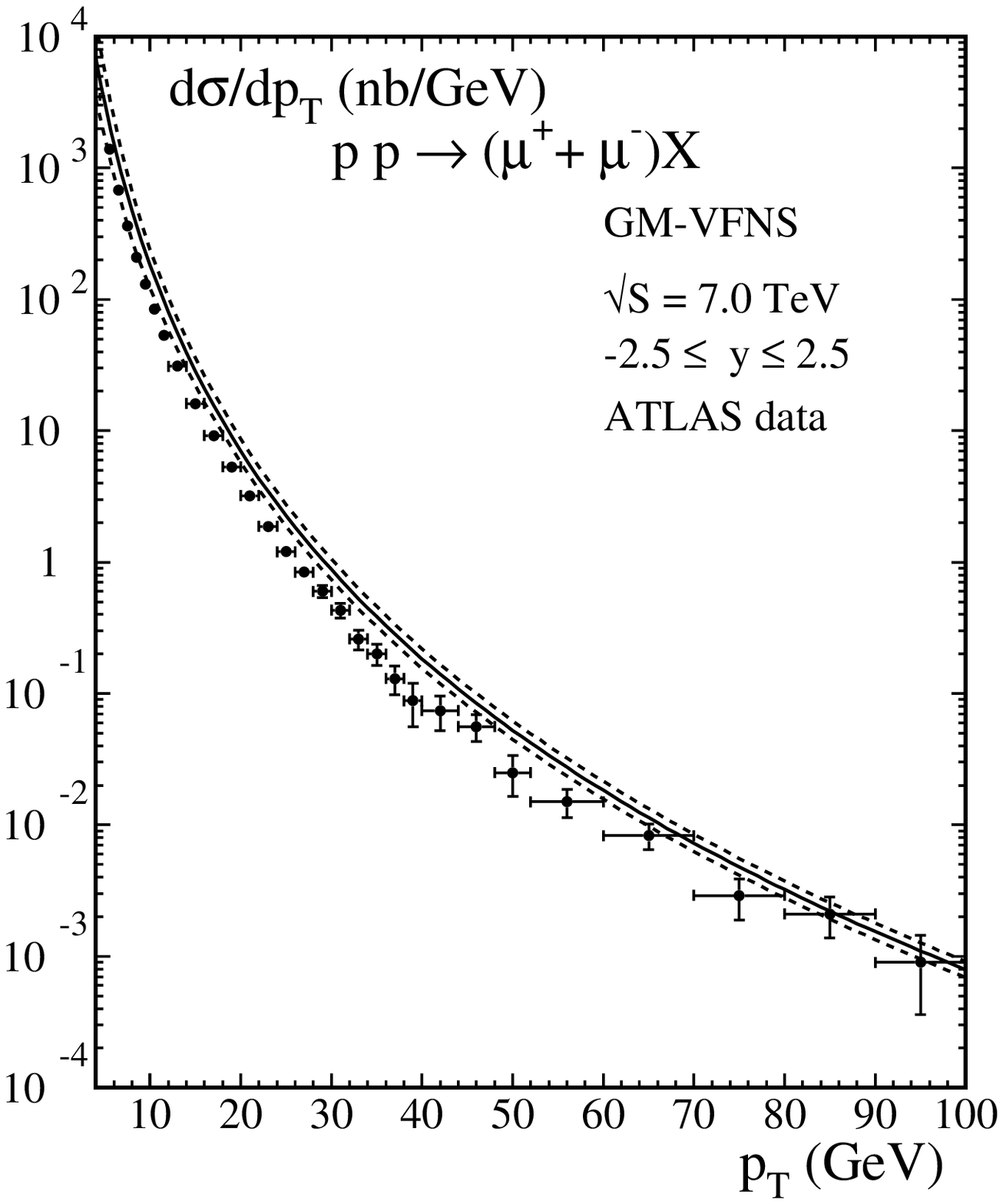}
\caption{\label{fig:fitgraphs3} GM-VFNS predictions for inclusive leptons from
charm and bottom hadron decay compared to ATLAS \cite{6} data.}
\end{figure*}
\begin{figure*}[ht]
\includegraphics[width=6cm]{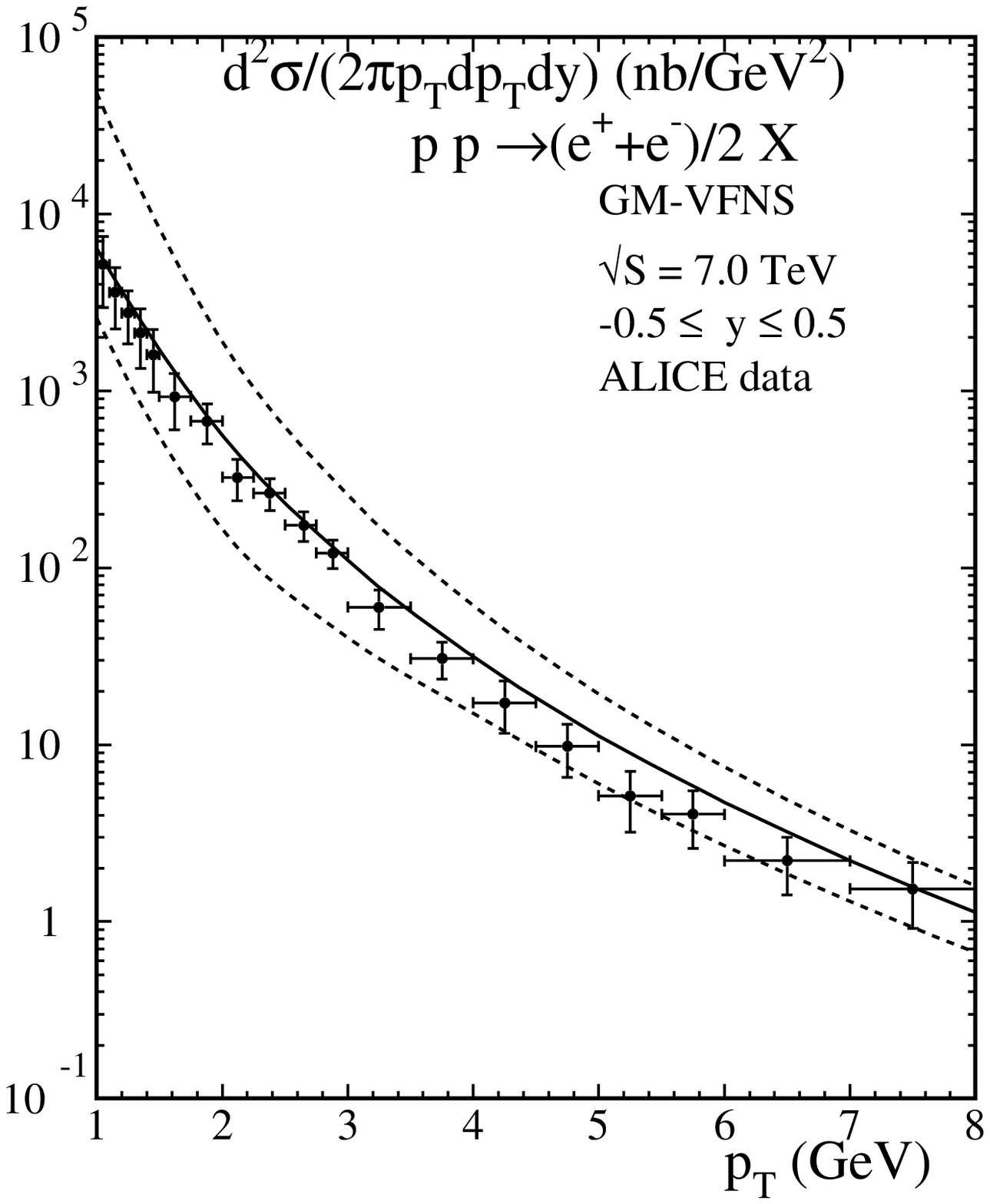}
\includegraphics[width=6cm]{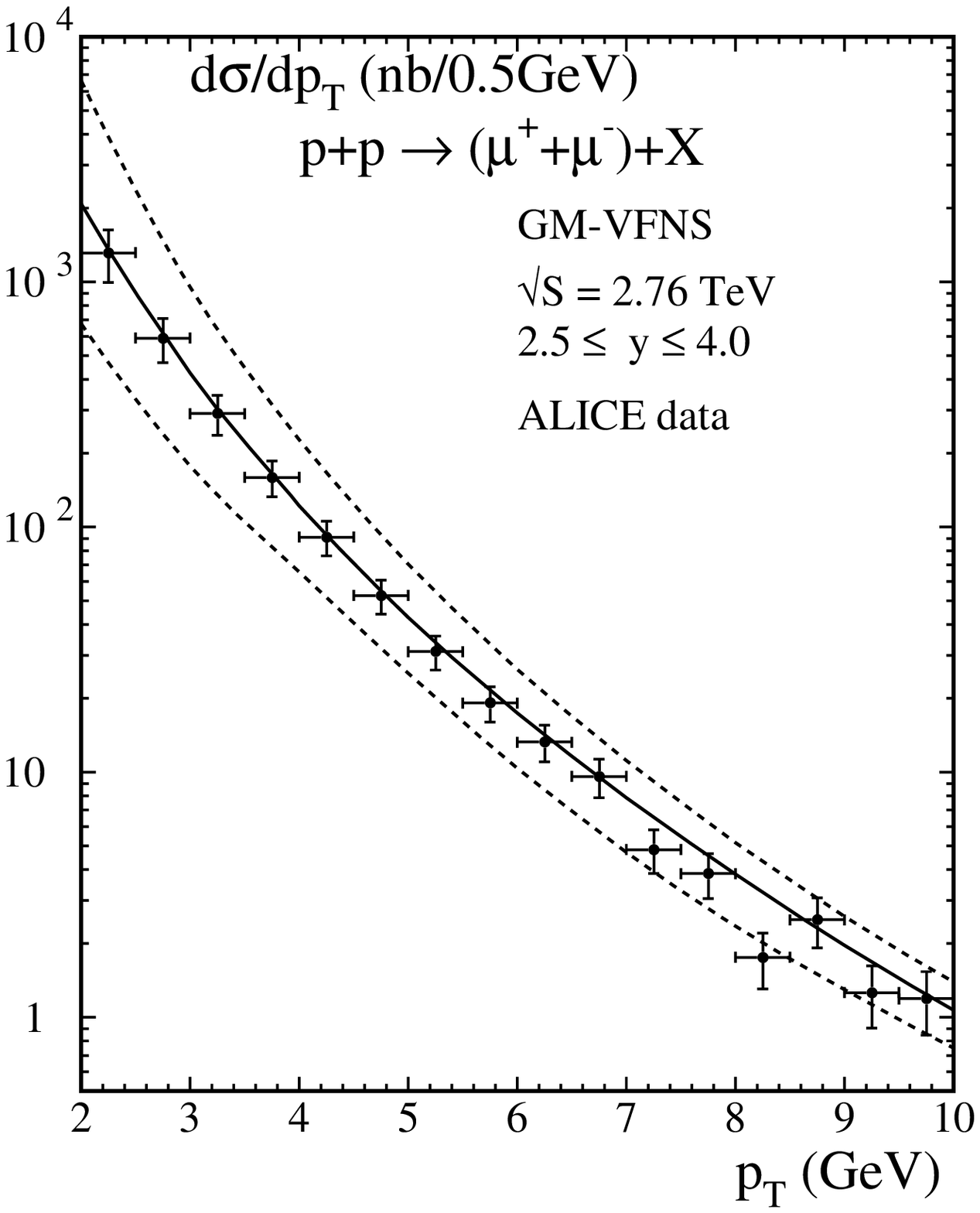}
\includegraphics[width=6cm]{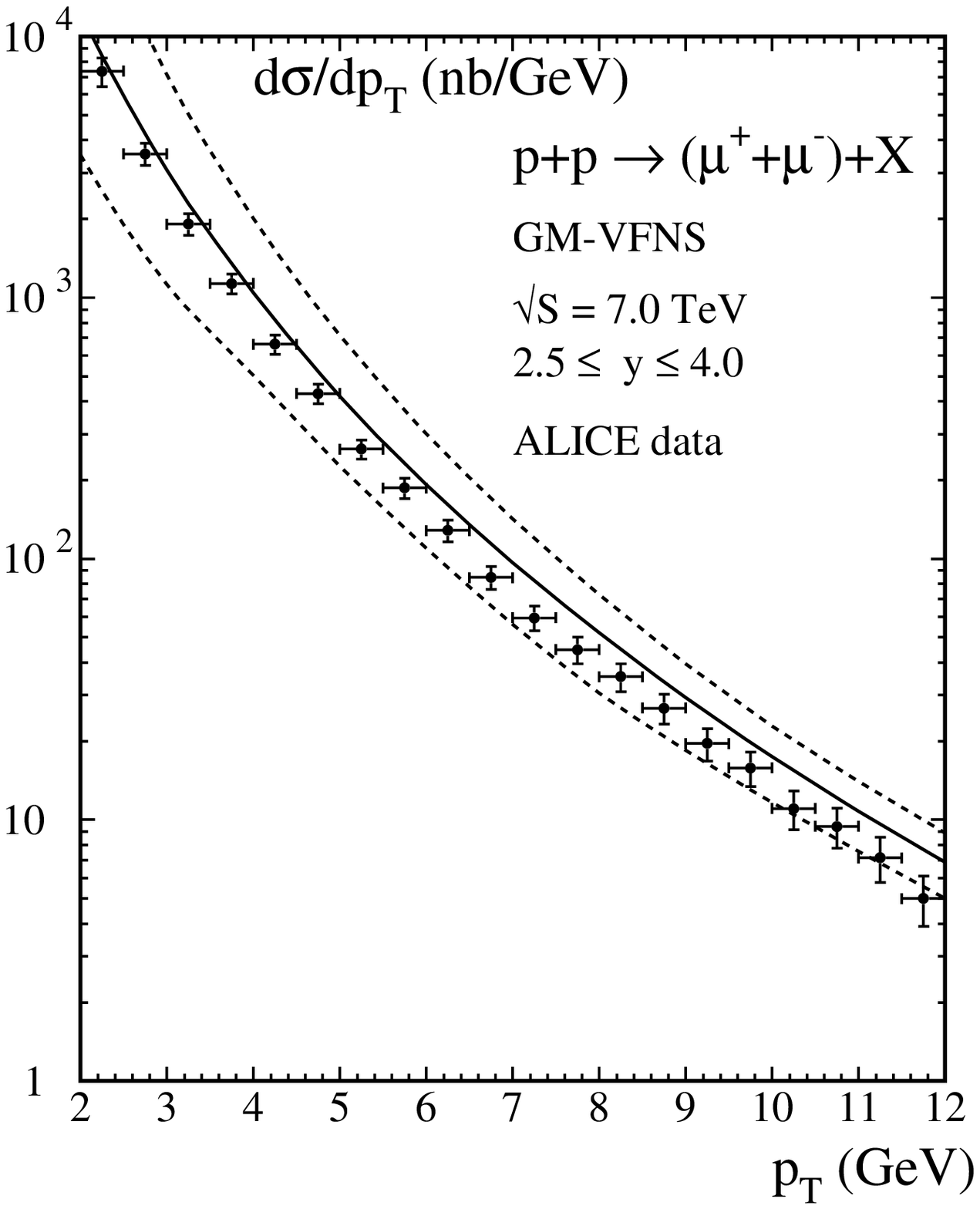}
\includegraphics[width=6cm]{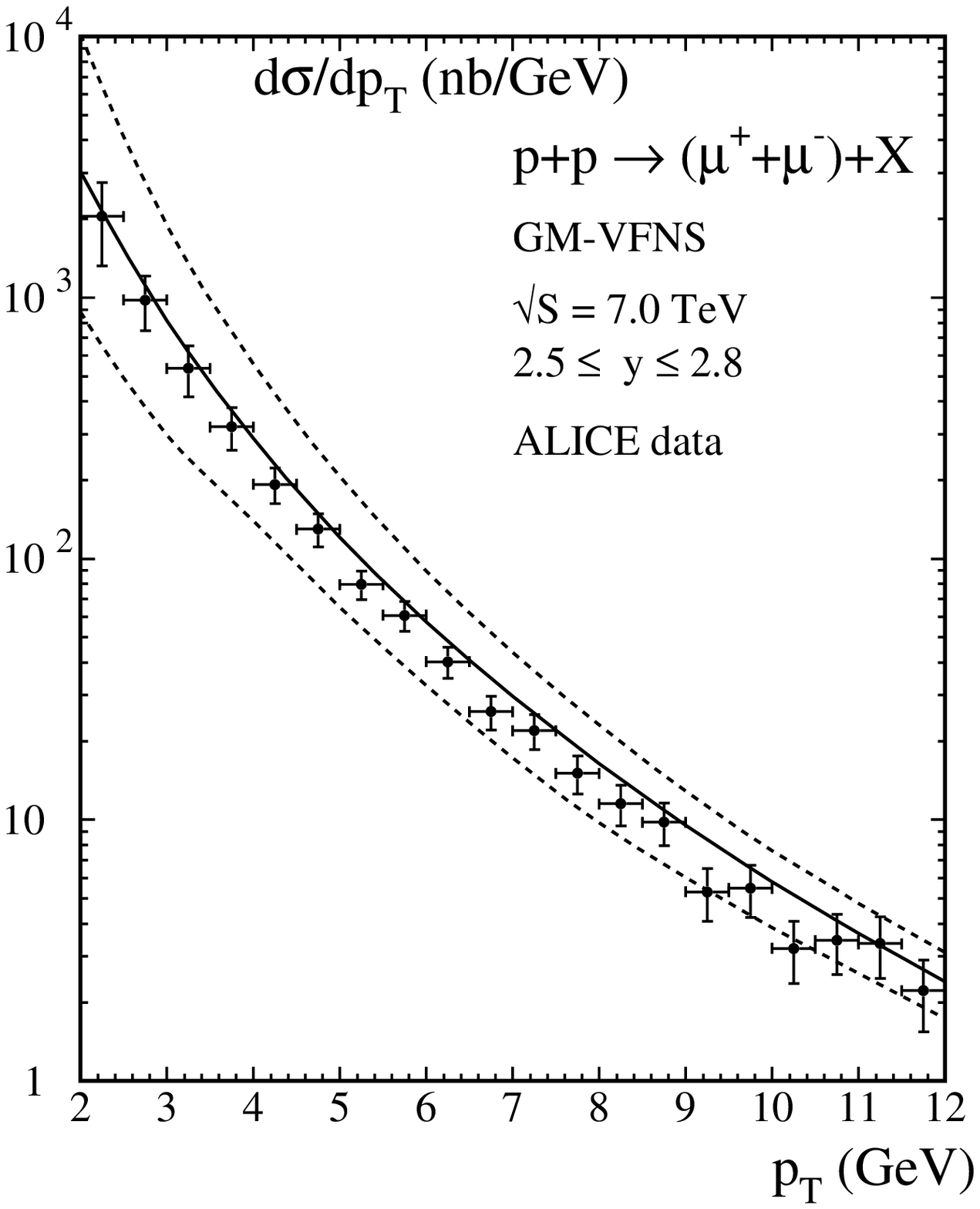}
\caption{\label{fig:fitgraphs4} GM-VFNS predictions for inclusive central and
forward lepton production from the sum of charm and bottom hadron decay 
compared to ALICE \cite{9,8,7} data.} 
\end{figure*}
\begin{figure*}[ht]
\includegraphics[width=6cm]{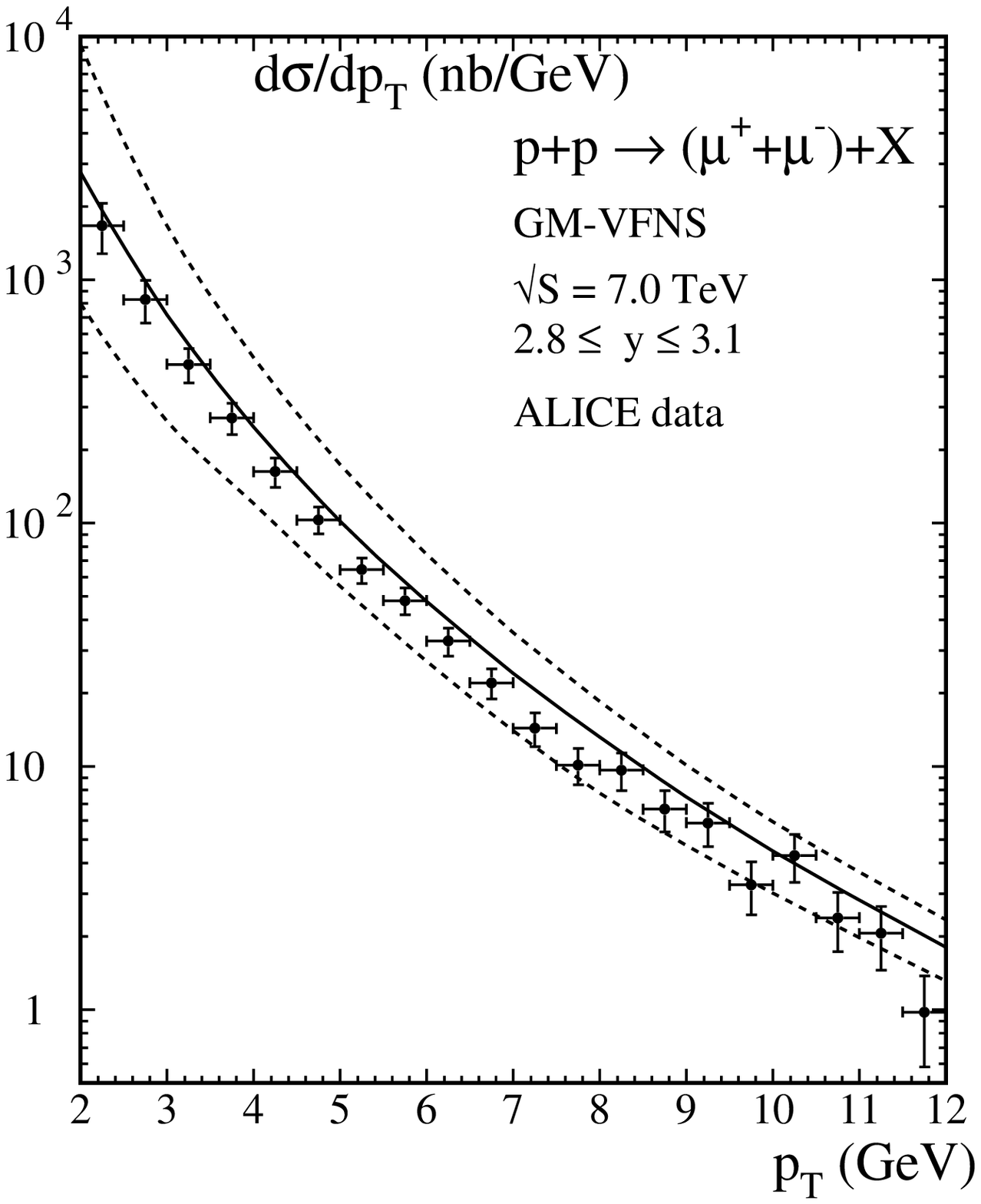}
\includegraphics[width=6cm]{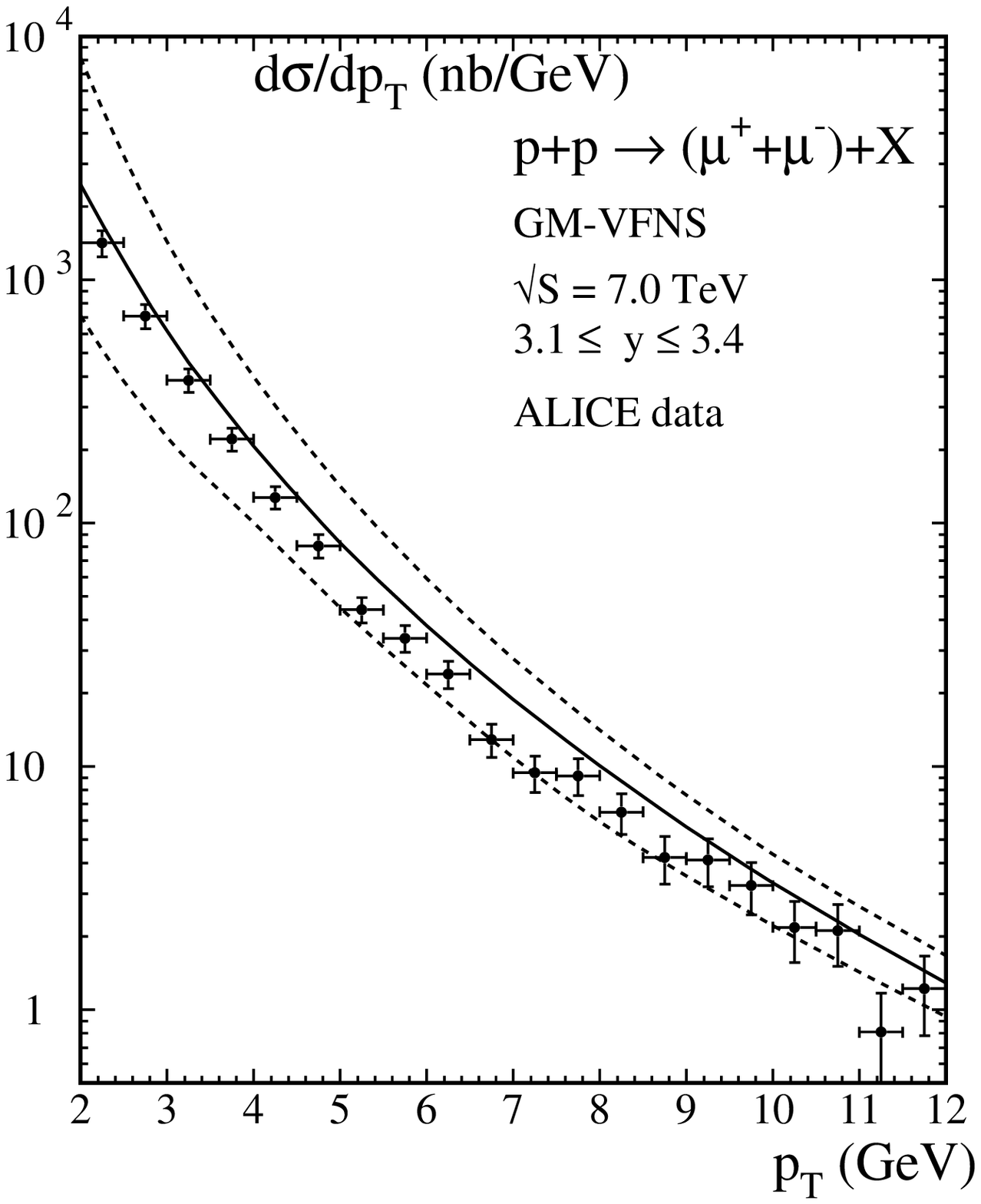}
\includegraphics[width=6cm]{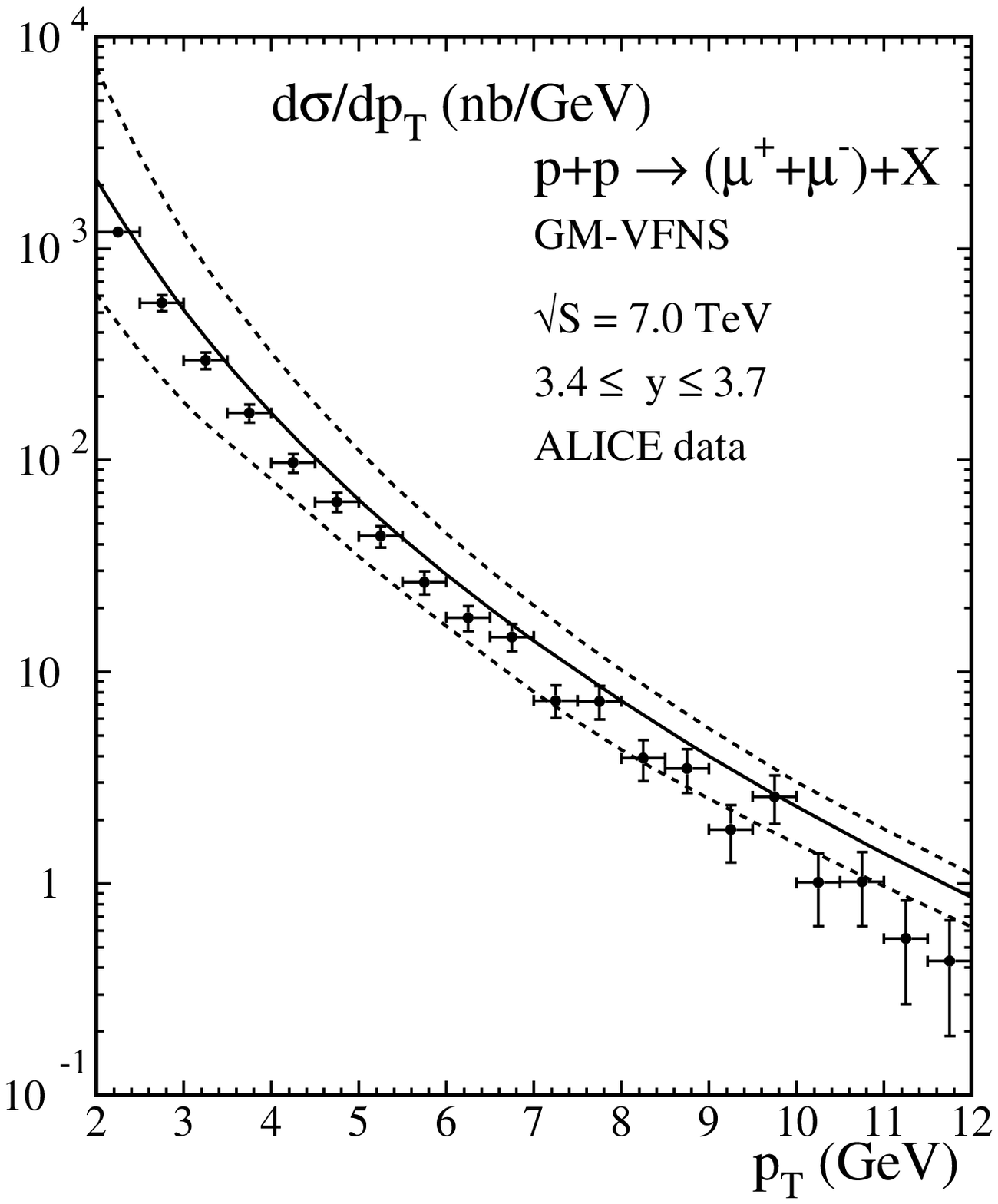}
\includegraphics[width=6cm]{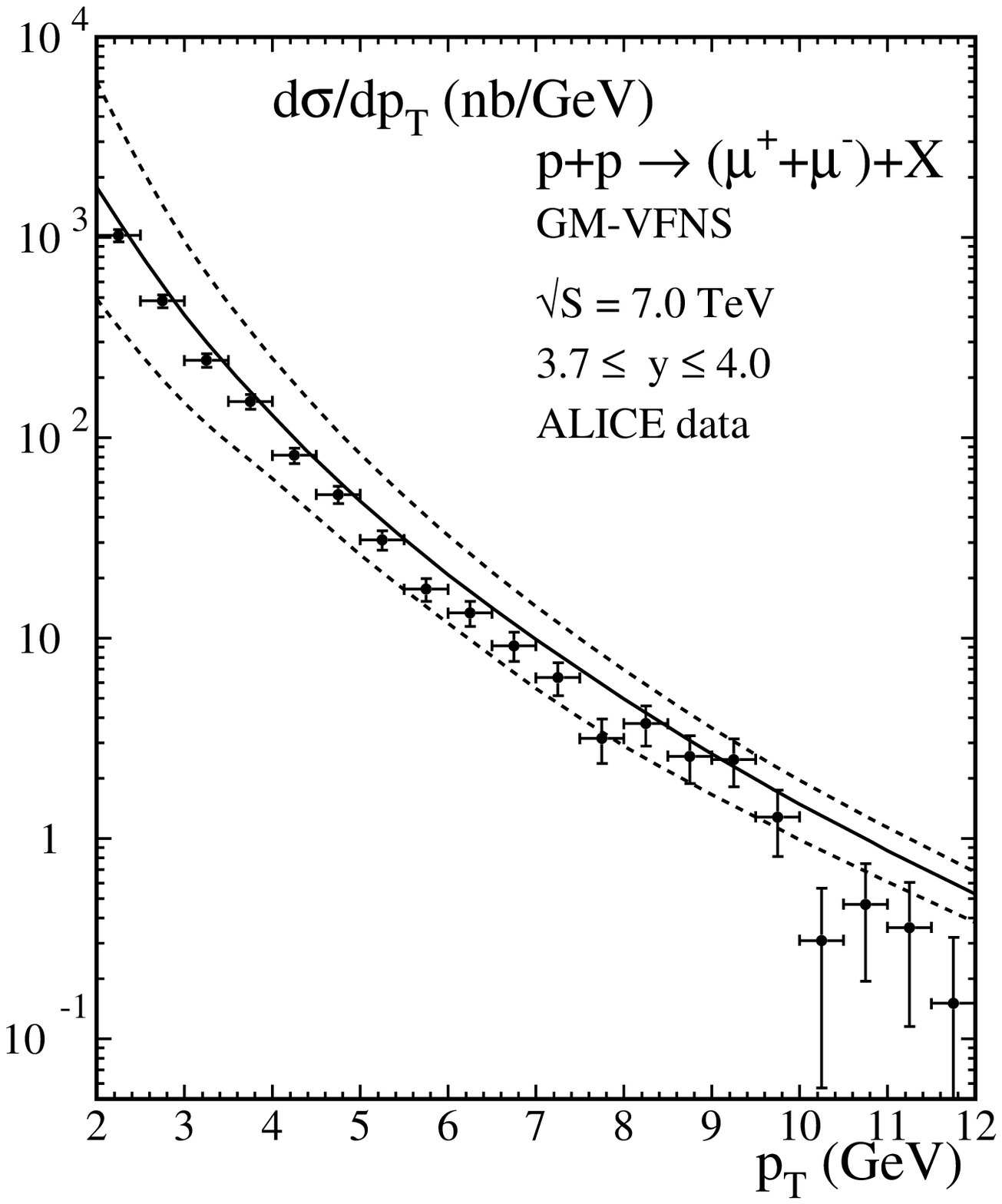}
\caption{\label{fig:fitgraphs5} GM-VFNS predictions for inclusive forward lepton 
production from charm and bottom hadron decay compared to ALICE \cite{7} data
in various rapidity regions.}
\end{figure*}

Next we present our predictions for the ATLAS and ALICE kinematical conditions
for the combined cross sections $d\sigma/dp_T$ for electron (or muon) 
production resulting from the sum of charmed and bottom hadron production for 
$\sqrt{S}=7$ TeV and one results from ALICE for $\sqrt{S}=2.76$ TeV. This 
latter measurement was done by the ALICE collaboration in order to have the
reference $pp$ cross section for their $PbPb$ measurements at the same energy.
In general the contributions from $D$ ($B$) decays dominate at small (large)
$p_T$ and are approximately equal at $p_T \simeq 4$ GeV.
The ATLAS data \cite{6} consist of three different measurements of 
$d\sigma/dp_T$ :(i) for the production of electrons ($e^++e^-$) as a function 
of $p_T$ for $7 \leq p_T \leq 26$ GeV integrated over the rapidity range 
$-2.0 \leq y \leq 2.0$ excluding the rapidity interval 
$1.37 \leq |y| \leq 1.52$, (ii) the production of muons ($\mu^++\mu^-$) in the 
same $p_T$ and $y$ range as in (i), and (iii) the production of muons 
($\mu^++\mu^-$) in the $p_T$ range $4 \leq p_T \leq 100$ GeV and the $y$ range 
$-2.5 \leq y \leq 2.5$. The result 
of our predictions is shown in Fig. \ref{fig:fitgraphs3}. 
The comparison of the data with the predictions is not so good. 
The data in all three figures are mostly below the predictions for 
the default scale choice.
The second group of data comes from the ALICE collaboration. These are cross
sections $d\sigma/dp_T$ for muon production at forward rapidity in the region 
$2.5 \leq y \leq 4.0$ at $\sqrt{S}=7$ TeV \cite{7} and at $\sqrt{S}=2.76$ TeV
\cite{8}. The data for $d\sigma/dp_T$ are presented as a function of $p_T$ 
beteen $2.0 \leq p_T \leq 12$ GeV, integrated 
over the total rapidity range and separately over five bins: (1) 
$2.5 \leq y \leq 2.8$, (2) $2.8 \leq y \leq 3.1$, (3) $3.1 \leq y \leq 3.4$,
(4) $3.4 \leq y \leq 3.7$ and (5) $3.7 \leq y \leq 4.0$ as a function of $p_T$  
in the same transverse momentum range . The data
and the comparison with our predictions are shown for the total $y$-range in 
Fig. \ref{fig:fitgraphs4} (second figure $\sqrt{S}=2.76$ and third figure $\sqrt{S}=7$ TeV) and 
for the cross sections integrated over the five
$y$ bins (1) to (5) in Fig. \ref{fig:fitgraphs4} (most right figure) and in  Fig. \ref{fig:fitgraphs5}. 
The 
agreement with the data is very satisfactory. The data points with their 
errors lie mostly between the curve for the default scale choice (full line) 
and the prediction for the minimal scale choice (lower 
dashed line). In addition
we show in Fig. \ref{fig:fitgraphs4} also the results for the later 
published ALICE cross sections
in the central rapidity region ($|y| \leq 0.5$) (most left figure). The
agreement of the ALICE data \cite{9} and our predictions with the default 
scale choice is again quite good.

\section{Summary}

We have calculated the cross sections for inclusive lepton production
originating from heavy flavour decays at LHC c.m. energies of 2.76
and 7 TeV in the framework of GM-VFN scheme and compared them with
measured cross sections $d\sigma/dp_T$ in different rapidity regions of the 
ALICE, ATLAS and CMS collaborations at the LHC. We found generally 
good agreement with
the experimental data inside the experimental and theoretical accuracies
for both, the lepton data coming only from bottom hadron decays (CMS and
ALICE data) and the lepton data for the sum of charm and bottom hadron
decays. This shows that within the considered kinematical regions as 
given by the
LHC collaborations the description of the inclusive production of charmed
hadron and bottom hadron is well accounted for. Except the CMS data all
other inclusive lepton production data from the LHC collaborations have been 
compared to predictions of the FONLL approach \cite{3bis,3} and similar good 
agreement between data and theoretical predictions has been found.

\section*{Acknowledgments}

We thank R. Maciula for sending us the data points of the BABAR collaboration. 
This work was supported in part by the German Federal Ministry
for Education and Research BMBF through Grant No.\ 05~H12GUE, 
by the German Research Foundation DFG through Grant No.\ 
KN~365/7--1, and by the Helmholtz Association HGF through 
Grant No.\ Ha~101.
%
%

\end{document}